# Microwave Current Imaging in Passive HTS Components by Low-Temperature Laser Scanning Microscopy (LTLSM)

A.P. Zhuravel[1], Steven M. Anlage[2] and A. V. Ustinov[3]

We have used the LTLSM technique for a spatially resolved investigation of the microwave transport properties, nonlinearities and material inhomogeneities in an operating coplanar waveguide $YBa_2Cu_3O_{7-\delta}$ (YBCO) microwave resonator on an $LaAlO_3$ (LAO) substrate. The influence of twin-domain blocks, in-plane rotated grains, and micro-cracks in the YBCO film on the nonuniform *rf* current distribution were measured with a micrometer-scale spatial resolution. The impact of the peaked edge currents and *rf* field penetration into weak links on the linear device performance were studied as well. The LTLSM capabilities and its future potential for non-destructive characterization of the microwave properties of superconducting circuits are discussed

## 1. INTRODUCTION

Thin-film devices made from high-$T_c$ superconductors (HTS) are encouraging candidates for the improvement of existing electronics [1-5]. Much attention has been paid to passive microwave devices, including delay lines, multiplexers, resonators and filters for mobile, cellular and satellite communications [6-9]. However, despite their expected excellent performance, a number of specific problems (of both fundamental and technological origin) remain unsolved for superconducting (SC) structures at radio frequencies (*rf*). The *global* (integral) electronic properties of the devices under varied external conditions are strongly dependent on *microscopically* distributed inhomogeneities of SC parameters and local defects, which can cause the SC device to perform below expectations [10--17]. Also, non-uniform current densities lead to nonlinear *rf* response of the devices even at modest microwave power [18-26]. Finally, the investigation of the *intrinsic* sources of degradation and nonlinearity at high frequencies is important for finding the limiting characteristics of such SC devices.

A powerful instrument for non-destructive spatially resolved characterization of SC devices is the Low-Temperature Laser Scanning Microscope (LTLSM). This allows *in-situ* (at cryogenic temperatures) 2-D imaging of distributed parameters in SC films as a function of the local critical current [27]. It is possible to manifest local resistive characterization of the SC transition [28-30], the spatial variations of magnetic and electromagnetic fields [31-34], and to carry out 2-D defectoscopy of structural and technological faults in operating SC devices and circuits [35-37]. In this work, we apply the LTLSM technique to analyze the redistribution of *rf* current flow in the vicinity of typical microdefects in YBCO films leading to anomalously high *rf* photoresponse, and possibly to nonlinear microwave behavior of SC devices.

## 2. EXPERIMENTAL

### 2.1. Samples

Measurements were performed on two HTS resonators designed for operation at a temperature of approximately 70 K. This co-planar wave-guide (CPW) resonator was fabricated from a 240 nm thick $YBa_2Cu_3O_{7-\delta}$ (YBCO) film deposited on a 500 $\mu$m thick LaAlO$_3$ (LAO) substrate by laser ablation, and has a center strip line of 500 $\mu$m width and 7.75 mm length. The line was separated from the ground planes by 650 $\mu$m and coupled to the feed lines via two capacitive gaps 500 $\mu$m wide. The 20x10x0.5 mm$^3$ sample chip was glued by vacuum grease to a brass microwave package. The resonator packaging is shown schematically in Fig. 1. Silver paste was applied to the Au contacts laser ablated onto the ground planes. Electrical contacts were made by spring loading the center pins of the SMA connectors to the corresponding Au contacts on the center conductor lines through a thin indium layer. The transition temperature of the YBCO film was measured to be 86 K. The fundamental resonant frequency $f_0$ for the CPW resonator at 78 K was approximately $f_0$ = 5.2 GHz with a loaded Q $\approx$ 2500.

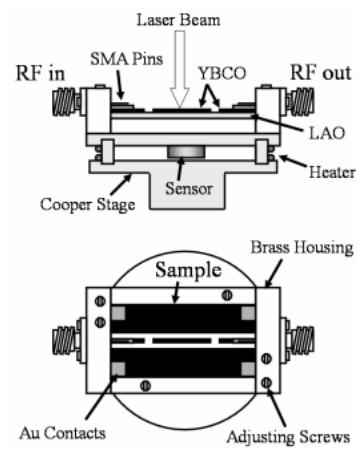

**Fig. 1.** The microwave package for CPW resonator

[1] B. I. Verkin Institute for Low Temperature Physics and Engineering, National Academy of Sciences of Ukraine, Kharkov, Ukraine
[2] Department of Physics, University of Maryland, College Park, MD, USA
[3] Physics Institute III, University of Erlangen-Nuremberg, Erlangen, Germany
Corr. author: Steven M. Anlage, Department of Physics, University of Maryland, College Park, Maryland, 20742-4111, USA, Tel: +001 301 405 7321, email: anlage@squid.umd.edu

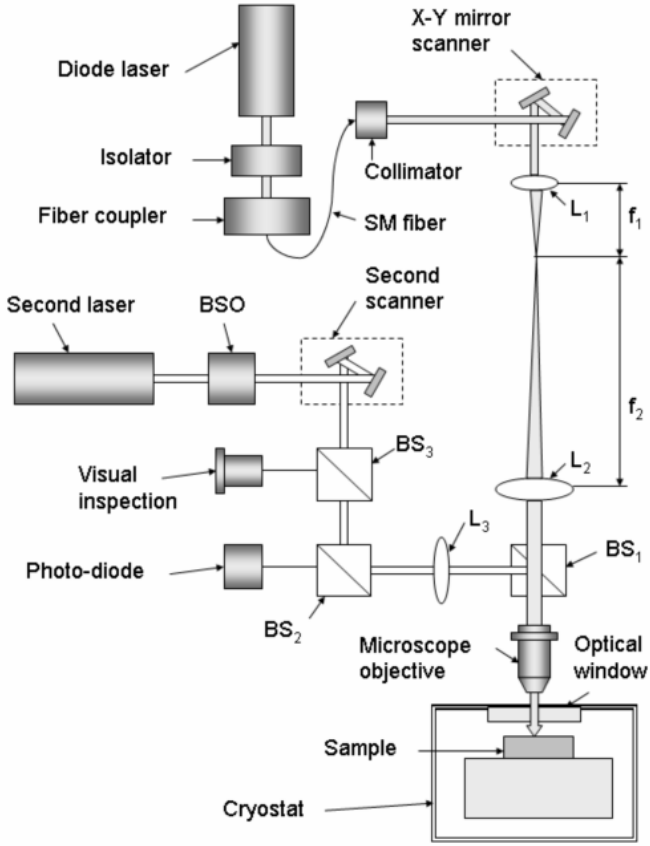

**Fig. 2**. Schematic diagram of the LTLSM optics and cryogenics. Notations: L - lens; BS - beam splitter; f - focal length; SM - single mode; BSO - beam-shaping optics.

**2.2. LTLSM technique**

Our LTLSM apparatus (see Fig. 2) consists of a typical reflected light-scanning microscope equipped with a miniature optical cryostat. This cryostat stabilized the temperature T of the sample in the range 78 - 300 K with an accuracy of 5 mK by using a resistive heater bifilarly coiled around a copper cold stage. The microwave package (see Fig. 1) was screwed to this stage inside a vacuum cavity of the cryostat and SMA coupled by stainless semirigid coaxial cables to the circuit delivering *rf* power $P_{RF}$. Under the operating condition, the CPW resonator was *x-y* scanned by a laser probe at different T and $P_{RF}$. To increase the signal-to-noise ratio of the LSM measurements, the amplitude of the laser intensity in the probe is typically modulated at a frequency of $f_m$ = 100 kHz by a TTL signal from the lock-in oscillator.

To form a Gaussian light probe on the sample surface, the laser beam travels from the diode laser (wavelength: $\lambda_{opt}$ = 670 nm, maximum output power: $P_{laser}$ = 10 mW), passed through an optical $\lambda/2 + \lambda/4$ isolator, and then through the fiber coupler to be spatially filtered in the $d_k$ = 4 $\mu$m core of a single-mode optical fiber with a numerical aperture (N.A.) of about 0.11. The transmitted laser radiation then is collimated to the size of the scanning mirrors, expanded by a factor $f_2/f_1 \approx 4$ by a pair of the relay lenses ($L_1$ and $L_2$), and passed through the beam splitter ($BS_1$) to fill the entrance pupil of the LSM objective lens. Three replaceable microscope objectives (20x, N.A. = 0.42; 5x, N.A. = 0.14 and 2x, f-theta) can be used, providing the opportunity for fast and accurate laser scanning of samples with a raster area ranging from 250x250 $\mu$m$^2$ up to 2.5x2.5 mm$^2$, with a spatial resolution from 1.2 to 6 $\mu$m, correspondingly. Taking into account absorption of the laser radiation by the optical components of the LSM system, the resulting intensity of the laser probe did not exceed $10^3$ - $10^5$ W/cm$^2$ on the sample surface. In order to scan the probe over the sample in a raster pattern, two orthogonal crossed mirrors, driven by a pair of galvano-scanners, are assembled in the plane conjugate to the entrance pupil of the objective lens.

Note that our LTLSM has several unique features that distinguish it from similar laser scanning techniques described in the literature. As seen in Fig. 2, this microscope is equipped with additional optics to organize a second scanning laser probe for local irradiation of the sample. For this purpose, a second laser ($\lambda_{opt}$ = 670 nm, $P_{laser}$ = 7 mW) with a structured beam is focused on the sample surface through the same objective lens. Beam-splitting ($BS_1$ - $BS_3$), scanning, as well as relay ($L_2$) optics of the second LSM channel have been created to independently focus and position both laser beams on the sample. The purpose of this independently scanning probe is to manipulate the thermo- and light- sensitive properties of HTS films (such as critical current density, magnetic penetration length and surface resistance) locally. This procedure enables one to artificially model faults, defects and inhomogeneities in microwave devices in order to study their influence on the *local* and *global* characteristics of the device.

Both (*probing and manipulating*) LSM channels together with an ordinary microscope visual inspection unit were rigidly fixed to the *x-y* translation table to be precisely electrical signal proportional to the local sample normal-incidence reflectivity as a function of probe position.

The second (*rf current imaging*) mode produces a map of local *rf* current density [32, 34, 38]. A synthesized signal generator (transmitted microwave power ≈ -15 to 0 dBm) excited a resonant mode of the CPW resonator. The absorbed laser power produces a thermal spot on the HTS film surface at any location *x, y* of the probe, oscillating in time with the frequency of laser beam modulation $f_m$. The thermal interaction of the probe with the sample shifted both $f_0$ and Q of the device due to changes in the local magnetic penetration depth and the stored energy in the resonator [32, 34, 39]. This caused a change in the $S_{12}(f)$ transmission curve that is proportional to the local *rf* current density squared $J_{RF}^2(x,y)$ at the location of the probe, and leads to a change in transmitted power P [see Fig. 3(b)]. A crystal diode detects the *rf* amplified changes in laser-modulated *rf* power, and an image of these changes $\delta V(x,y) \approx J_{RF}^2(x,y)$ is recorded and rescaled to be proportional to the $J_{RF}(x,y)$ amplitude, and then used for plotting the spatio-amplitude 3-D map of the $J_{RF}(x,y)$ distribution.

In the third (*IMD imaging*) mode, two fixed frequency signals ($f_1$ and $f_2$) were applied to the resonators [see. Fig. 3(a)]. The $f_1$ and $f_2$ were centered on the $|S_{12}(f)|$ curve with a close spacing from 10 kHz to 1 MHz and had the same amplitudes. Changes in $P_{f1}$ or $P_{2f1-f2}$ as a function of position *(x,y)* of the laser beam perturbation on the sample were imaged. A spectrum analyzer was used to measure the power in the tones ($P_f$) at these intermodulation frequencies to see exactly where in the device these tones are generated. Presumably, the more nonlinear parts of the material will contribute a bigger change to the IMD power when they are heated. Based on numerical simulations with the two-fluid model, we can assume to first approximation that the contrast seen by the LSM tuned to the intermodulation frequency is proportional to the local change in intermodulation current density scale, $J_{IMD}$ [40].

### 2.4. Spatial limitations

The resolution S of the LTLSM instrument in reflective imaging mode is limited by the size of the beam probe. Thermal diffusion in the sample also worsens the LTLSM contrast in *rf* PR and IMD PR imaging modes. Roughly, the smallest features of scanned LSM images may be resolved on a length-scale not less than $S = (d_{opt}^2 + l_T^2)^{1/2}$, where $d_{opt}$ is the optical resolution [in terms of the full-width at half-maximum (FWHM) of the focused Gaussian laser beam], and $l_T$ is the thermal healing length. As a test to determine the contribution of each component to S independently, the sharp features of the PR at the edge of a patterned YBCO film on LAO substrate was imaged using different LTLSM detection modes.

Figure 4(a) shows the variation (open circles 1) of local reflectance R(y) along a 50 μm y-line scan, measured by using a 20x LSM objective lens with 0.2 μm steps, through the YBCO/LAO interface. The Gaussian fit (solid moved relative to the cryostat. The general area of interest on the sample can be chosen by stepper motors in the range as large as 25x25 mm$^2$ with accuracy as small as 0.5 μm.

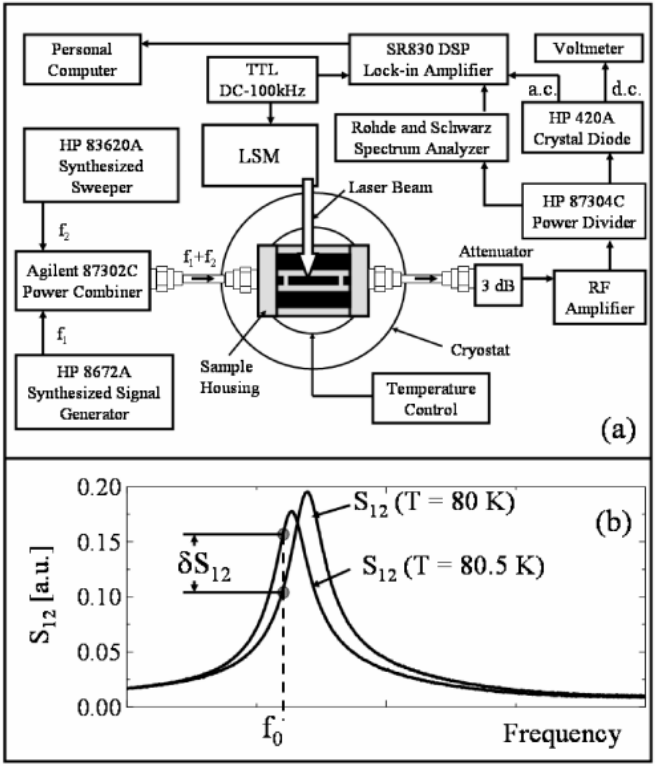

**Fig. 3.** (a) Schematic diagram of LTLSM electronics used for $J_{RF}(x,y)$ and IMD imaging, and (b) the principle of operation of the microscope.

### 2.3. Basic principles of LTLSM imaging

The principle of LTLSM operation is to scan the surface of a thin-film SC device with a tiny-focused laser beam (probe) for 2-D reconstruction of the response signal $\delta V(x,y)$ arising from the laser-excited area of the sample. The SC device acts as a self-detector and the theoretical challenge is to construct a physical model leading from the localized irradiation to the mechanism of the detected photo-response (PR). Fig. 3 depicts the measuring electronics incorporated in the LTLSM setup (see Fig. 2) for the PR processing.

A scanning light probe reaches the sample surface, which may absorb some of the light and reflect some of the light. Depending on which dominant PR mechanism is LTLSM detected, three different mapping modes were employed for imaging.

First, *the optical LSM images* were acquired in reflective contrast LSM mode by using a photo-diode detector of the returned light. It was helpful to establish the correlation between local microwave properties of the resonators with its topographic features in the same scale of spatial resolution and from the same area. In this mode, the modulated laser beam is reflected from the sample surface and monitored by an optical sensor to produce an *ac*

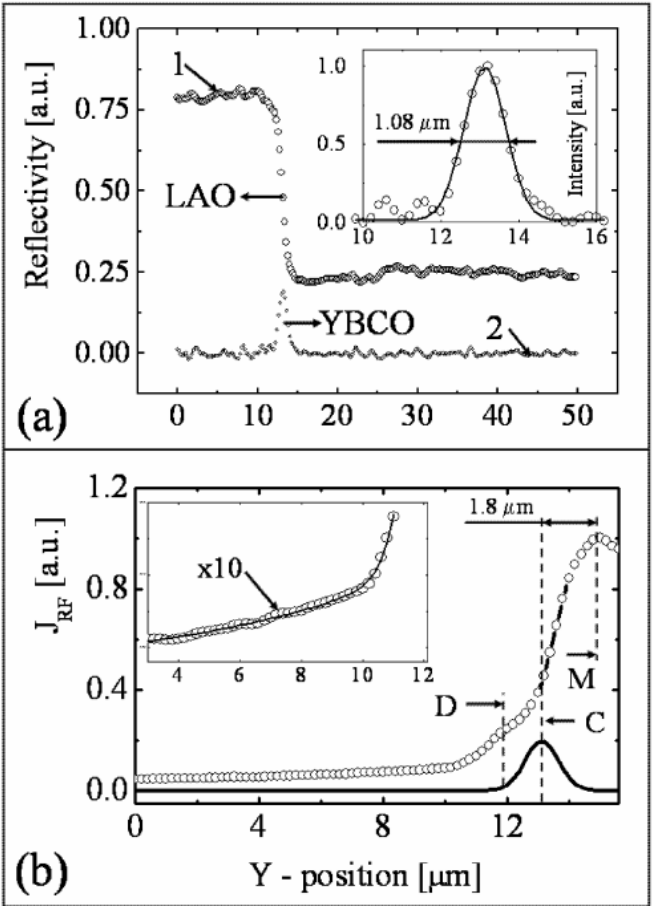

**Fig. 4**. LTLSM profiles of (a) reflective and (b) *rf* photoresponse. Details are described in the text.

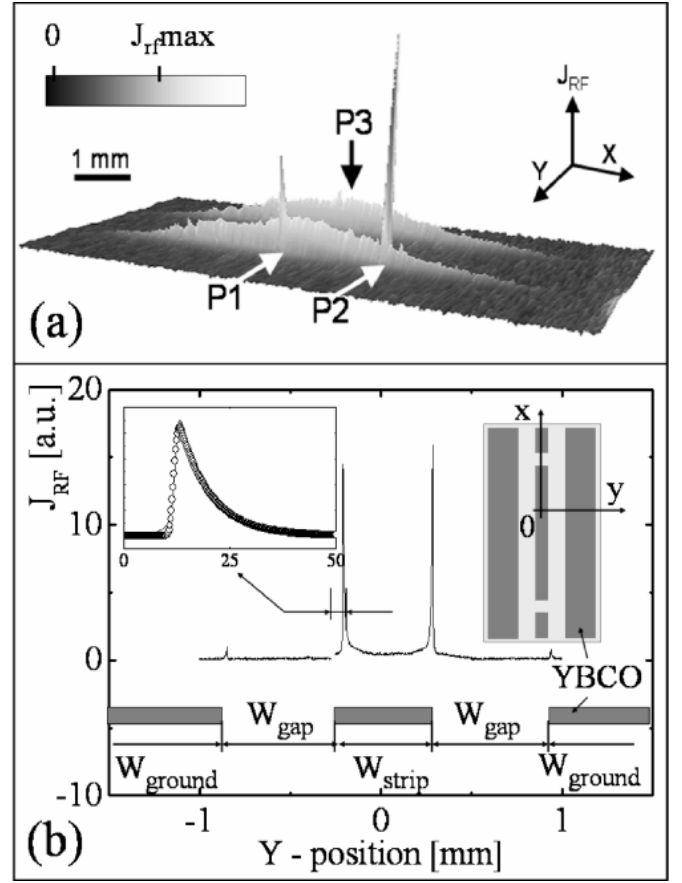

**Fig. 5**. LTLSM plots showing *rf* current distribution at T = 79 K in a YBCO film on LAO substrate (a) over a 1 x 8 mm$^2$ central area of the CPW resonator including the 0.5 x 7.75 mm$^2$ strip-line conductor and (b) along a single *y*-scan across the CPW mid-section

line in the inset) of the R(y) derivative (open diamonds 2) gives $d_{opt}$ ~ FWHM = 1.08 $\mu$m. The small disagreement between this value and the limiting size of the calculated diffraction spot (0.97 $\mu$m) is due to the geometric roughness of the YBCO strip edge. As is evident from Fig. 4(a), the peak position of the R(y) derivative precisely defines the location of the HTS strip edge. This fit curve is repeated in Fig. 4(b) (lowest curve) for use as a control curve to explain the features of the *rf* PR along the same line scan. The upper curve (diamonds) in Fig. 4(b) depicts a detailed profile of *rf* PR that is rescaled to be proportional to the local *rf* current density $J_{RF}$.

The local features of $J_{RF}(y)$ demonstrate that the LSM *rf* PR near the strip edge is described by different mechanisms of laser-sample interaction. In the region lying between lines D and M, the increasing *rf* PR corresponds to light- and thermo- induced modulation of $J_{RF}$ by absorbed energy from the oscillating laser probe. The *rf* PR is peaked inside the HTS strip at a distance 1.8 $\mu$m (line M) away from the strip edge (line C). This is a result of the convolution between the exact position of the $J_{RF}$ peak at the edge, and the bolometric collection of PR by a thermal spot with radius equal to the thermal healing length $l_T$. To the right of M, the *rf* PR decays in conformity with the falling Meissner current toward the strip center. To the left of D, heating of the LAO substrate by the probe is contributing to the LSM response formation. In this case, the strip edge plays the role of a micro-bolometer which detects a hot-spot from the laser probe. This feature can be used to determine $l_T$ experimentally.

The inset in Fig. 4(b) shows an enlarged profile of LSM PR (circles) together with an exponential fit (solid line). The value of $l_T$ ~ 4 $\mu$m extracted from this fit is in good agreement with values in the literature [41]. Hence our estimate of the spatial resolution of the LTLSM is S = 4.2 $\mu$m at a laser intensity modulation rate $f_m$ = 100 kHz. As mentioned above, the resolution in our setup depends on the selected scanning field and objective lens. Increasing the scanning field can be done at cost of having poorer spatial resolution.

### 3. RESULTS AND DISCUSSION

Figure 5(a) represents a 3-D LTLSM image showing the spatial variation of the *rf* current density along the CPW central strip-line at its resonance mode. The image was acquired at T = 79 K, $P_{RF}$ = -10 dBm and at a frequency that is 3 dB below the fundamental resonance peak [$f_0$ = 5.2 GHz, see Fig. 3(b)].

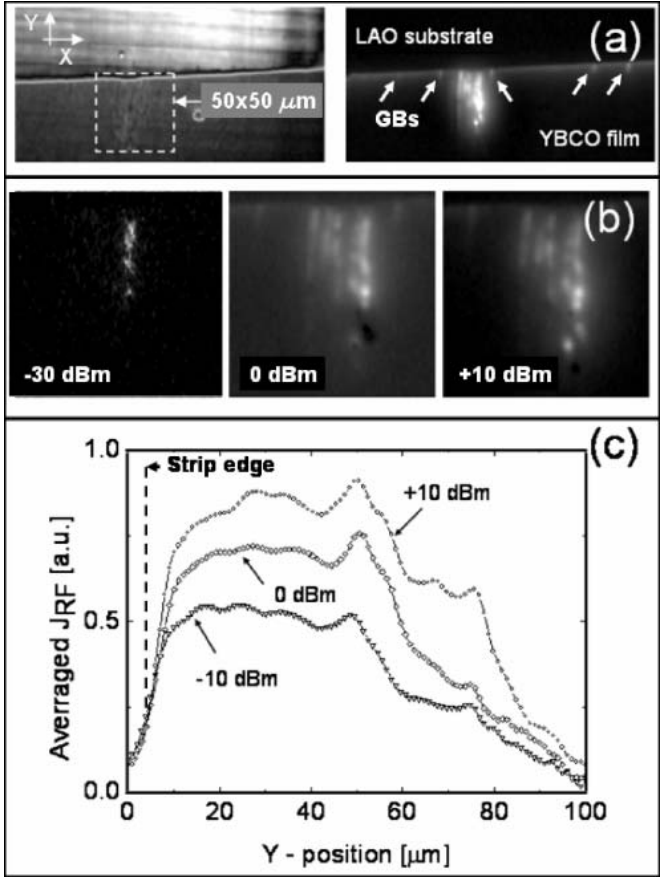

**Fig. 6.** (a) 200 x 100 $\mu m^2$ LTLSM images showing optical (upper) and *rf* (lower) properties of the YBCO film, including an in-plane rotated grain (position P1), and (b) *rf* power dependence of averaged *y*-scans showing detailed $J_{RF}$ penetration into the YBCO film near P1. The averaging was carried out within the LSM images (c) obtained in a selected 50 x 50 $\mu m^2$ zone of the sample at T = 80 K.

As evident from the plot, the $J_{RF}(x,y)$ has a typical $\cos(kx+\varphi)$ standing wave pattern along the *x* direction, reaching a maximum at the mid cross-section of the CPW resonator [38, 42]. Almost all the $J_{RF}(x,y)$ spatial profiles look like that shown in Fig. 5(b): a flux-free Meissner state having the two peak lines nearly localized at both strip edges [26, 43-45]. Exceptions are seen in a few areas marked by the arrows P1, P2 and P3 in Fig 5(a), where anomalously high photoresponse amplitude was seen. The possible reasons for such *rf* current pattern distortion were studied in detail.

Figure 6(a) shows LTLSM maps of both reflectivity and of $J_{RF}(x,y)$ distribution of the same area, scanned at T = 80 K, $P_{RF}$ = 0 dBm near the point P1 in Fig. 6(a) and Fig. 5(a). In this area, the anomalously-high LTLSM response is illustrated in $J_{RF}(x,y)$ imaging mode at all experimentally applied $P_{RF}$ in the range from -30 dBm to +10 dBm. The first measured $J_{RF}(x,y)$ distribution appears at -30 dBm [see left image in Fig. 6(b)] as an individual line almost perpendicular to the YBCO strip edge. A few more lines of the same orientation gradually filled in the closed area of the HTS film when $P_{RF}$ was increased from -20 to +10 dBm [Fig. 6(b)]. Fig. 6(c) shows the power dependence of $J_{RF}(x,y)$ redistribution in the area of maximum LTLSM *rf* photoresponse. To get these profiles at different *rf* powers, we averaged $J_{RF}(x,y)$ data in a 50x50 $\mu m^2$ area of the HTS strip. The area for the averaging is shown by the dotted box in Fig. 6(a). It is evident that there is a deep *rf* field penetration into the film that increases with increasing $P_{RF}$.

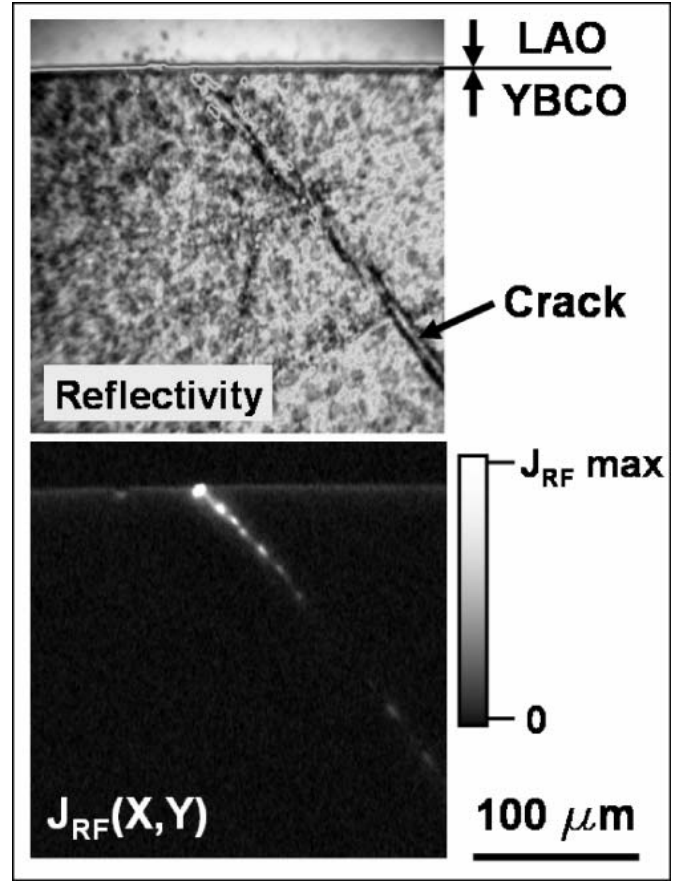

**Fig. 7.** Gray-scale representation of 250 x 250 $\mu m^2$ LTLSM images that were simultaneously obtained in reflective (upper) and $J_{RF}$ (lower) modes. Inhomogeneous $J_{RF}(x,y)$ distribution shows a deep penetration of microwave field in the HTS strip. Spatial modulation of *rf* current densities along the crack may be formed by localized vortex pinning on a twin-domain structure of the YBCO film.

By comparison between the $J_{RF}(x,y)$ map and the reflectivity LSM image, we established that the anomalously-high LSM PR in this area is due to the influence of weak links within an individual YBCO grain having different microwave properties from the rest of the film. The distribution of $J_{RF}(x,y)$ inside this grain is dictated by the direction of twin-domain blocks which are orthogonally oriented to those in the neighboring grain areas. In addition, the reflectivity of this grain was established to be a little higher as compared with its neighbors. This may be due to oxygen-depletion of this grain [46], or of its a-axis orientation. [47].

Also, points P2 and P3 in Fig. 5(a) were LTLSM analyzed. Nontrivial behavior of both the ordinary *rf* and IMD PR was detected near P2 in the vicinity of a crack in the YBCO film. The crack [see Fig. 7] is formed by an area of sharp misorientation of twin domain blocks in LAO [40].

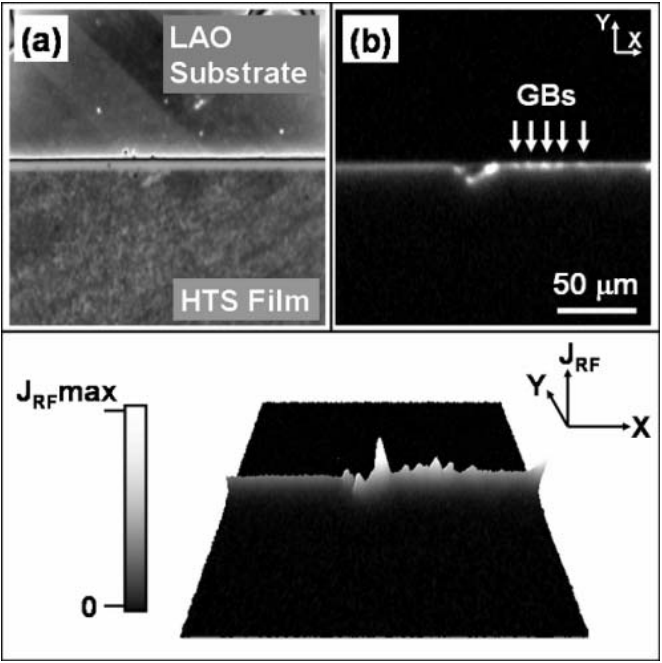

**Fig. 8.** Detailed view of the YBCO film on LAO substrate showing (a) reflectivity, (b) 2D- and (c) 3D- photo-response of the $J_{RF}(x,y)$ in a 200 $\mu$m x 200 $\mu$m$^2$ area. The substrate clearly shows the linear patterns of twin domain blocks crossing the patterned edge of the YBCO thin film. The patterned edge is crenellated on the length scale of the twin domain blocks, as well as due to inhomogeneous associated with the grain boundaries (GBs) produced by substrate twinning.

Inhomogeneous distribution of $J_{RF}(x,y)$ here shows a deep penetration of microwave field in the HTS strip. Spatial modulation of *rf* current densities along the crack may be formed by localized vortices pinned on a twin-domain structure of the YBCO film.

Additionally, a meandering of the current due to irregularities associated with the substrate twin domain blocks in the etching of the YBCO film was found at P3. Figure 8 demonstrates a one-to-one correspondence between edges of the film disrupted by the underlying twinned substrate and significant enhancement of the $J_{RF}(x,y)$. Both causes of irregularity of the distribution of $J_{RF}(x,y)$ along the YBCO strip edge due to irregularities (meandering) in the film etching [38] and due to inhomogeneities associated with the individual grain boundaries (GBs) produced by substrate twinning are presented in the figure.

These irregularities, together with the individual misoriented grains having nontrivial *rf* properties, are paramount candidates for enhanced nonlinear response from the device. However, even if the task of repairing these faults is accomplished, another extrinsic effect still dominants the nonlinear response of the CPW resonator, masking its intrinsic limit associated with the *d*-wave pairing state in superconductors [21].

Weak links resulting from grain boundaries in the HTS material are dominating the lower limit of nonlinearities [1]. The influence of weak links on the edge current density of the resonator is clearly visible in Figures 6(a) and 8(b) under the white arrows. These weak links are grain boundaries produced by large-scale twin domain blocks in the substrate crossing the HTS strip edge. The weak links have a lower critical supercurrent density, and this is accompanied by an increase of the effective magnetic penetration length into the strip. To confirm this fact experimentally, we carried out a model LSM demonstration using the second laser beam available in our microscope.

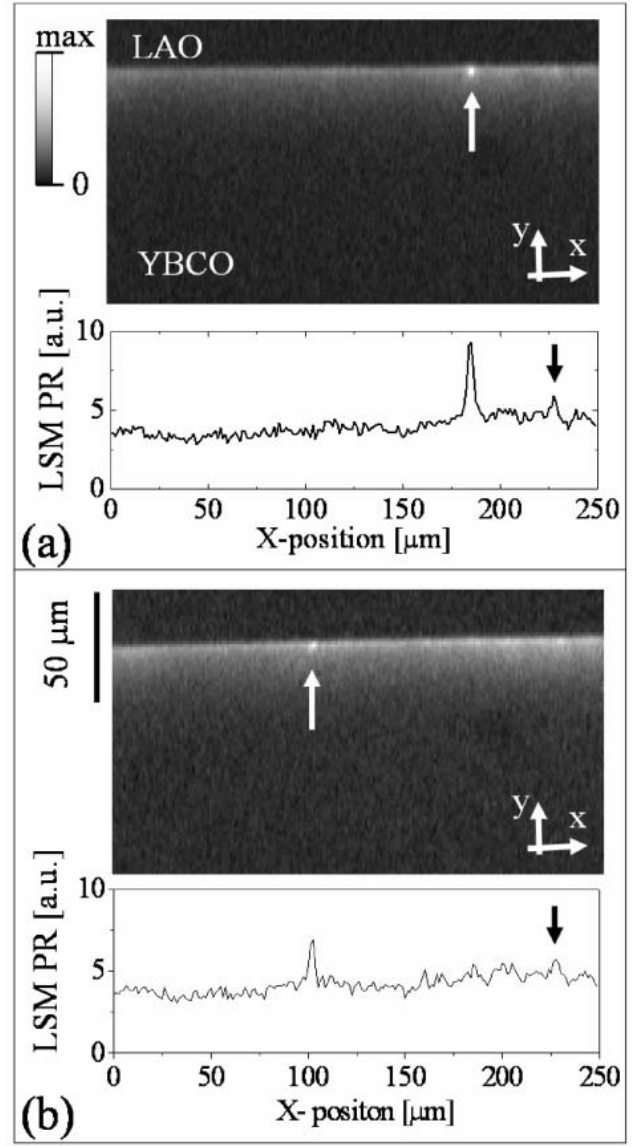

**Fig. 9**. LTLSM images in *rf* PR mode showing a spatial modification of the $J_{RF}$ distribution in the YBCO strip under the influence of a second laser probe positioned along the strip edge at T = 80 K. The white arrow indicates the location of the second beam while the black arrows indicate positions of natural grain boundaries at the edge of the film.

Figure 9 displays the $J_{RF}(x,y)$ distributions in the same 250x125 $\mu$m$^2$ area of the CPW resonator at two fixed positions of an additional continuous wave (*cw*) LSM laser probe focused at the edge of the YBCO strip. A flaw-free film edge was chosen for LTLSM imaging in *rf* current mapping mode at T = 80 K and $P_{RF}$ = 0 dBm. The white arrows in Fig. 9 indicate the different positions of the second laser probe. Those points look brighter in the grey-scale LTLSM pictures and show up as peaks in graphs showing the amplitude of *rf* photoresponse along the edge. The enhanced *rf* PR at the location of a laser beam focus is associated with thermally-induced changes in local penetration depth caused by the *cw* laser heating. The visible difference in amplitudes of the *rf* PR peaks in Fig. 9(a) and (b) is caused by variations in laser heating because the probe is positioned 2 $\mu$m deeper into the film in Fig. 9(a) compared to Fig. 9(b), where the laser focus is exactly at the edge. The far-field influence of overheating caused by the additional laser probe is detected by natural grain boundaries indicated in Fig. 9 by black arrows. It is evident that the amplitude changes in *rf* PR there is governed by thermally-induced changes in $\lambda$. These grain boundaries also produce enhanced nonlinear response from the device, and will be investigated in detail in the future.

## 4. CONCLUSION

This paper demonstrates some advantages of the LTLSM technique for spatially resolved failure analysis of superconducting microwave devices. We have demonstrated that a typical superconducting microwave device has at least four different classes of defects and irregularities that can potentially cause enhanced nonlinear response. We also present the first preliminary results on two-beam LTLSM imaging, demonstrating that local in-situ manipulation by *rf* properties in operating microwave devices is possible simultaneously with a LSM mapping.

**ACKNOWLEDGMENTS**

We thank K. Harshavardhan of Neocera, Inc. for providing the CPW resonator samples. We acknowledge the support of a NATO Collaborative Linkage Grant PST.CLG.977312, NSF/GOALI DMR-0201261, and the Maryland/Rutgers/NSF MRSEC DMR-0080008 through the Microwave Microscope Shared Experimental Facility.Final Stage.